\documentclass[11pt]{article}
\usepackage[margin=1in]{geometry}

\usepackage{amsmath, amssymb, amsthm}
\usepackage{proof}
\usepackage{stmaryrd}
\usepackage{url} 

\newcommand{\lfprove}[3]{#1 \vdash #2 \mbox{:} #3}

\newcommand{\arr}[2]{#1 \rightarrow #2}
\newcommand{\type}{\mbox{\sl Type}}
\newcommand{\typedlam}[3]{\lambda #1:#2.\, #3}
\newcommand{\typedpi}[3]{\Pi #1\mbox{:}#2.\, #3}

\newcommand{\app}{\ }
\newcommand{\lfobj}{\mbox{\sl tm}}
\newcommand{\lftype}{\mbox{\sl ty}}
\newcommand{\hastype}{\mbox{\sl hastype}}
\newcommand{\istype}{\mbox{\sl istype}}
\newcommand{\allx}[2]{\forall_{#2} #1.}

\newcommand{\enc}[1]{\llbracket #1 \rrbracket}

\newcommand{\natty}{\mbox{\sl nat}}
\newcommand{\ztm}{\mbox{\sl z}}
\newcommand{\numty}{\mbox{\sl num}}

\title{On Encoding LF in a 
  Predicate \\ Logic over Simply-Typed Lambda Terms}
\author{Gopalan Nadathur \and Mary Southern}

\begin{document}
\maketitle

\begin{abstract}
Felty and Miller have described what they claim to be a faithful
encoding of the dependently typed $\lambda$-calculus LF in the logic of
hereditary Harrop formulas, a sublogic of an intuitionistic variant of
Church's Simple Theory of Types. 
Their encoding is based roughly on translating object expressions in
LF into terms in a simply typed $\lambda$-calculus by erasing
dependencies in typing and then recapturing the erased dependencies
through the use of predicates.
Unfortunately, this idea does not quite work.
In particular, we provide a counterexample to the claim that the
described encoding is faithful. 
The underlying reason for the falsity of the claim is that the mapping
from dependently typed $\lambda$-terms to simply typed ones is not
one-to-one and hence the inverse transformation is ambiguous.
This observation has a broad implication for other related encodings.
\end{abstract}

A faithful encoding of a dependently typed $\lambda$-calculus within
a predicate logic can be useful for a variety or reasons: 
it can help us understand the relative expressive power of the different
systems, it can provide a means for implementing type checking in
the $\lambda$-calculus, and it can be the basis for a framework for
reasoning about specifications developed in the $\lambda$-calculus.
An encoding that is motivated by such considerations has been provided
by Felty and Miller~\cite{felty90cade} for the Edinburgh Logical Framework
(LF)~\cite{harper93jacm} in the logic of hereditary Harrop
formulas~\cite{miller12proghol}.
This encoding has been claimed to be faithful: specifically, this
claim is the content of Theorem 5.2 in~\cite{felty90cade}.
Unfortunately, the mentioned theorem is false.
We show this to be the case in this note by providing a
counterexample to it.

Our counterexample is based on an LF signature that includes the following
type-level constructors: $\natty:\type$ and
$\numty:\typedpi{x}{\natty}{\type}$.
Further, the signature includes the following object-level constants:
$\ztm:\natty$ and
$c:\typedpi{w}{(\typedpi{x}{\natty}{\typedpi{y}{(\numty \app x)}{\natty}})}
           {\natty}$. 
Given this signature, we may construct the object-level expression
$(c \app (\typedlam{x}{\natty}{\typedlam{y}{\numty~\ztm}{\ztm}}))$.
Our counterexample focuses on the typeability of this expression.
More specifically, it considers the derivability of the LF judgement
$\lfprove{\cdot}
         {(c \app (\typedlam{x}{\natty}{\typedlam{y}{\numty~\ztm}{\ztm}}))}
         {\natty}$.
It is easily seen that this judgement is in fact not derivable in LF
but, as we show below, the encoding of Felty and Miller leads to a
different conclusion.

The encoding of LF derivability questions in the logic of hereditary
Harrop formulas works in three steps.
In the first step, LF type and object expressions are translated into
terms of two distinguished types in a simply typed
$\lambda$-calculus.
Specifically, the type $\lftype$ is used for type expressions and
$\lfobj$ is used for object expressions.
To support this translation, the LF signature must be reflected into a
suitable signature in the target language.
In the example under consideration, this results in a signature with
the following constants: $\natty : \lftype$,
$\numty:\arr{\lfobj}{\lftype}$, $\ztm : \lfobj$, and
$c:\arr{(\arr{\lfobj}{\arr{\lfobj}{\lfobj}})}{\lfobj}$.
Using this signature, the LF object-level expression under
consideration would be represented by the simpy-typed $\lambda$-term
$(c \app (\typedlam{x}{\lfobj}{\typedlam{y}{\lfobj}{\ztm}}))$.

The translation carried out in the first step loses important typing
information; in particular information about dependencies in typing is
erased.\footnote{The specific translation described loses more
  information than just dependencies in typing since it collapses all
  object types into the single type $\lfobj$.
  However, the counterexample we
  present will remain one even under a translation that avoids this
  defect.}
The second step of the encoding tries to capture the lost information
through the use of a unary predicate \istype\ that is intended to
identify encodings of well-formed types and a binary predicate
\hastype\ that relates encodings of object expressions to those of
(dependent) types. 
The realization of this step requires the LF signature to be
translated into a collection of formulas that define these two
predicates.
Focusing only on the \hastype\ predicate, something that suffices for
our counterexample, this step yields the following formulas for
this predicate in the context under consideration:
\begin{tabbing}
\qquad\=$\allx{w}{\arr{\lfobj}{\arr{\lfobj}{\lfobj}}}($\=\qquad\qquad\=\qquad\=\kill
\>$\hastype\app \ztm\app \natty$, and\\
\>$\allx{w}{\arr{\lfobj}{\arr{\lfobj}{\lfobj}}}
           (\allx{x}{\lfobj}\hastype\app x \app \natty \supset$\\
\>\>       $\allx{y}{\lfobj}\hastype\app y\app (\numty\app x) \supset
                           \hastype\app (w\app x\app y)\app \natty)$ \\
\>\>\>$\supset \hastype\app (c\app w)\app \natty$
\end{tabbing}
In what follows, we will use the notation of \cite{felty90cade} in
denoting the collection of formulas obtained by translating the
signature $\Sigma$ in our example by $\enc{\Sigma}$.

The last step in the encoding consists of posing the validity of an LF
typing judgement as the derivability of a translated form of the
typing judgement from the formulas obtained from translating the
signature in the logic of hereditary Harrop formulas.
In the particular situation under consideration, this reduces to
considering the derivability of the formula
\begin{tabbing}
\qquad\=\kill
\>$\hastype\app (c\app (\typedlam{x}{\lfobj}{\typedlam{y}{\lfobj}{z}})) \app \natty$
\end{tabbing}
from the assumption formulas $\enc{\Sigma}$.

We now have all the pieces in place for our counterexample.
It is easy to see that the formula identified by the translation is in
fact derivable from $\enc{\Sigma}$, contrary to the earlier
observation that the LF judgement that it is supposed to encode is not
derivable.
More specifically, this example indicates that Theorem 5.2 in
\cite{felty90cade} is false in the ``if'' direction: derivability in
the encoded version \emph{does not} imply derivability in LF.
The reason for the mismatch should also be evident: the many-to-one
nature of the encoding allows us to conclude that \emph{some} LF
typing judgement from which the translated version is obtained is
valid but not that the specific LF judgement that is of interest is
valid. 
A closer examination of the counterexample allows us to trace the
problem even more specifically to the fact that some of the dependency
information that is lost in the transformation of LF object
expressions to simply typed $\lambda$-terms is not recovered by the
predicate-level encoding.

\section*{Acknowledgements}

The observations in this note were made in the course of conducting
research supported by the National Science Foundation under Grant 
No. CCF-1617771. 
Any opinions, findings, and conclusions or recommendations expressed
in this material are those of the authors and do not necessarily
reflect the views of the National Science Foundation.

\end{document}